\documentclass{raa}

\usepackage[T1]{fontenc} 

\usepackage{natbib}
\usepackage{cancel}
\usepackage{enumitem}    
        
\usepackage{amsmath} 	   
\usepackage{amssymb}       
\usepackage{amsfonts}
\usepackage{graphicx} 
\usepackage{dblfloatfix}	    
\usepackage{verbatim}
\usepackage{epstopdf}
\usepackage{tensor}
\usepackage{epsfig,multicol,bbm}
\usepackage{color}
\usepackage{colortbl}
\usepackage{float}
\usepackage{multirow}
\usepackage{array}
\usepackage{nicematrix}
\usepackage[colorlinks,linkcolor=red,anchorcolor=green,citecolor=blue]{hyperref}

\begin{document}
\title{Toward a direct measurement of the cosmic acceleration: The pilot observation of H I 21cm absorption line at FAST}
 
 \volnopage{Vol.0 (20xx) No.0, 000--000}      
 \setcounter{page}{1} 
 
 \author{Jiangang Kang
 	\inst{1,2}
 	\and Chang-Zhi Lu
 	 \inst{1, 2}
 	\and Tong-Jie Zhang
 	 \inst{1, 2}
 	\and Ming Zhu
 	  \inst{3}
}
 
  \institute{Institute for Frontiers in Astronomy and Astrophysics, Beĳing Normal University, Beĳing 102206, China; {\it tjzhang@bnu.edu.cn} \\
 	 \and Department of Astronomy, Beijing Normal University,  Beijing 100875, China \\
   	\and National Astronomical Observatories, Chinese Academy of ciences,Beijing  100101, China
}

\abstract {This study presents results  on detecting neutral atomic hydrogen (HI) 21cm absorption in the spectrum of PKS1413+135 at redshift $z=0.24670041$. The observation was conducted by FAST, with a spectral resolution of 10 Hz, using 10 minutes of observing time. The global spectral profile is examined by modeling the absorption line using a single Gaussian function with a resolution of 10 kHz within a 2 MHz bandwidth. The goal is to determine the rate of the latest cosmic acceleration by directly measuring redshift evolution of H I 21 cm absorption line with Hubble flow towards a same background Quasar over a decade or longer time span. This will serve as a detectable signal generated by the accelerated expansion of the Universe at redshift $z < 1$, referred  to as redshift drift $\dot{z}$ or the SL effect. The measured HI gas column density in this DLA system is approximately equivalent to the initial observation value, considering uncertainties of the spin temperature of a spiral host galaxy. The high signal-to-noise ratio of 57, obtained at a 10 kHz resolution, strongly supports the feasibility of using the H I 21 cm absorption line in DLA systems to accurately measure the redshift drift  rate at a precision level of around $10^{-10}$ per decade.    
\keywords{cosmology: redshift drift -- cosmology: dark energy-- galaxies: DLA-- radio: H I 21cm line} 
}

\authorrunning{J.G  Kang  et al.}            
\titlerunning{Measurement of the cosmic acceleration }  %

\maketitle
\section{Introduction}

   The occurrence that the accelerated expansion of the Universe since $\sim$ 5 Gyr ago was discovered \citep{1998AJ....116.1009R,1999PhRvL..83..670P}. It clearly indicates the existence of new physics beyond the Standard Model. Later on, the acceleration is extensively described by a phenomenological model, called 'dark energy', with the unusual equation of state parameter $\rm w_{DE} = p_{DE} /\rho_{DE} < -1/3$, $\rm w_{DE}$,$\rm p_{DE} $ and $\rm  \rho_{DE}$ respectively denotes  the state of equation , the pressure and energy density of dark energy. The past two decades,  the massive   observational experiments are devoted  to digest the evolution law of $\rm w_{DE}$  with redshift by a variety of cosmological probes. Until now, most of evidences seem  to point that the paranormal  resulted from the simplest form of dark energy, i.e. the cosmological constant $\Lambda$ ,corresponding to $\rm w_{DE} = -1$, finally shaping the  standard $\Lambda$CDM  cosmology  \citep{2023arXiv230714802E,2024ExA....57....5M}. High precision measuring the expansion history with the various observational tools  persists  a foremost subject in modern cosmology. Initially proposed by Sandage, the method to measure redshift variations in extragalactic objects was later refined by Loeb using Lyman-$\alpha$ forests in distant quasars. This observational technique is known to the Sandage-Loeb (SL) effect \citep{  1962ApJ...136..319S,1998ApJ...499L.111L}.
   
    The redshift drift effect( $\Delta z /\Delta t$ ,hereafter $\dot{z}$), as   the Cosmic Accelerometer, provides a high-precision method for directly observing the change in redshift of distant sources caused by the universe’s expansion, which offers a model-independent probe in cosmology\citep{Cooke_2019,2019BAAS...51g.137E,2022arXiv220305924C,2012PhRvD..86l3001M}. The effect is negligible, with redshift and velocity drift for a typical galaxy at redshift $z = 1$ being close to $10^{-10}$ for redshift drift and 3 $\rm cms^{-1}$ for velocity drift per decade respectively\citep{2015aska.confE..27K}, similar to gravitational accelerations in galaxies and clusters\citep{2008PhLB..660...81A,Darling_2012}. However, the tiny signal, which is purely radial and cosmological, can be identified without interference from the local motions of the observer\citep{Darling_2012}. It is  can be immediately distinguished from  the polarized drifts of proper acceleration\citep{Zakamska_2005,Titov_2011,2012SCPMA..55..329X}.  An alternative explanation for the redshift drift interference is that the apparent acceleration, similar to stars circling a galaxy's core, can be consistently discounted regardless of its strength or direction. Therefore, the overall impact of proper motion within a gravitationally bound system on redshift drift is negligible\citep{2022PDU....3701088L,1998ApJ...499L.111L,2008PhLB..660...81A,1986ApL....25..139T,1982ApL....22..123P}.

  Damped Lyman-$\alpha$ Absorber (DLA) systems, progenitors of spiral galaxies, have a characteristic H I column density of $\rm N_{H I} \ge 2 \times 10^{20}/cm^2$\citep{1986ApJS...61..249W,2015A&A...575A..44G,2001A&A...369...42K,Kanekar_2001,Gupta_2013}. A great amount of DLA systems presumably will contribute to the measurement precision of SL signal and lower the  error of velocity drift $\sigma_{\mathrm{V}}$. There are three main  differences between intervening and associated absorptions what concluded from \citep{2016MNRAS.462.4197C}: (i) That the mean associated profile is wider than the mean intervening profile;(ii) From a simple model of the H I column density distribution,
that the high velocity wings often observed in associated absorption, arised from the sub-pc gas, which appears to be absent in the
intervening absorbers, due to the associated absorption arises in AGN (radio galaxies and quasars), where the majority of gas that accreted into the central supermassive
black hole, but intervening absorption lines seem in more quiescent galaxies;
(iii) The consistency in the mean intervening profile widths to
either side of z $\sim$ 1, indicates no kinematical or thermal evolution with redshift.

 The majority of cataloged DLA  systems  are at redshifts $z \geq 2$, detected by the Sloan Digital Sky Survey (SDSS), because ultraviolet Lyman-$\alpha$ absorption lines at redshifts $z < 1.65$ are not observable with ground-based telescopes\citep{2004PASP..116..622P}, yet only around 50  DLAs in redshift $z \lesssim 1.7$ \citep{2021MNRAS.503..985A,2009MNRAS.396..385K}. Using the H I 21 cm absorption line from DLA systems as an alternative to the Lyman-$\alpha$ line  can measure redshift drift at any distance. This method also reveals more detailed insights into the early state of neutral gas in host galaxies. In the interstellar medium, neutral hydrogen exists in two stable phases: the Cold Neutral Medium (CNM) with a spin temperature of roughly 80-100 K, and the Warm Neutral Medium (WNM) at a much hotter spin temperature of about 5000-8000 K,while the temperature lies in $\rm 500 < T_s < 5000$ K as a unstable phase  hosts the substantial portion of neutral ISM. Actually the majority of the observed result of  gas spin temperature is a column density-weighted harmonic average from the mixed phase ISM\citep{2001A&A...369...42K,2023PASA...40...46K,Kanekar_2001}. 
 
 Redshifted 21 cm absorption lines in the spectra from strong radio sources can serve as a direct method to measure redshift drift\citep{Darling_2012,2015A&A...575A..44G,Yu_2014,Gupta_2013,2001A&A...369...42K,1992ApJ...400L..13C}.The European Extremely Large Telescope (ELT), a 40-meter class observatory equipped with a high-resolution spectrograph, aims to detect redshift signals between z=2 and z=5. By detecting the Lyman-alpha forests in the southern sky over a 20-year period\citep{2021arXiv211012242M,Liske_2008,2013arXiv1310.3163M,2021Msngr.182...27M}. 
 The forthcoming Square Kilometre Array (SKA) is set to achieve the goal
  by observing the H I 21cm line emissions from galaxies and the absorption lines from DLA systems across a redshift range of approximately 1 to 13\citep{2015aska.confE..27K,2004NewAR..48.1259K, 2019MNRAS.488.3607A,2020EPJC...80..304L,2023MNRAS.518.2853R,2021arXiv211012242M,2019arXiv190704495B,2015aska.confE.134M,2015aska.confE.167S,2015aska.confE..17A,marques2023watching}, meanwhile the CHIME \citep{Yu_2014,Newburgh_2014,Bandura_2014} has  performed this experiment at intermediate redshifts.  In addition, it is hopeful that FAST also can detect  this signal\citep{NAN_2011,8331324,2019SCPMA..6259502J}. By using a combination of blind searches and targeted observations at a redshift of $ z \le 0.35$, it is  expected to identify approximately 70 DLA  systems by scanning the celestial equator for one month. Extending the observing time to one year will yield around 800 DLA systems, five years could result in 1900, and a decade-long search can obtain the detections of 2600 DLAs\citep{Jiao_2020,2023A&A...675A..40H}.

  In this study, we report the detection of the H I 21 cm absorption line from the DLA system at a redshift of $z=0.24671 \pm 0.00001$ towards the quasar PKS1413+135. This observation was made using the ON-OFF mode of FAST telescope for a duration of approximately 10 minutes in November 2019\citep{2022PDU....3701088L}. Exploring the flat cosmological model includes  parameters such as $\rm H_0$, $\Omega_m$, $\Omega_{\rm DE}$, and the equation of state for dark energy, $w$. This object is categorized as a BL Lac or red quasar due to its compact polarized continuum emission and relativistic jet in its line of sight\citep{2023A&A...671A..43C}. The host galaxy, a Sb-c type spiral galaxy, exhibits rapid changes in centimeter-scale optical emissions, particularly intense at near-infrared wavelength\citep{1991MNRAS.249..742M},which is a young radio source with a compact symmetric object (CSO)  within 0.01" from the center in H-band imaging, showing a two-sided parsec-scale jet that is slightly bent, not aligned with the line of sight\citep{1996AJ....111.1839P},Moreover, Einstein's X-ray observation reveals a high H I column density of $\rm N_{H I} > 2 \times 10^{22}$ $\rm cm^{-2}$ and significant extinction with $\rm A_v > 30$ in soft X-ray data, indicating a gradual linear decline towards the IR spectrum\citep{1992ApJ...400L..17S,2023A&A...671A..43C,1996AJ....111.1839P}.

    Assuming that the spin temperature  of the H I gas cloud $\rm T_{spin} = 300$ K for the PKS1314+135 in this work since  its host galaxy  resembles a normal spiral  like Milky Way with the spin temperature  from 250  to  400 K across the disc\citep{1992ApJ...400L..13C,2009ApJ...693.1250D}. The final task of the investigations of H I 21 cm absorption lines  to detect the redshift drift signal  with FAST. These outcomes can be used to assert the rate of cosmic acceleration and rule out or place constraints on the competing candicates of dark energy and even furnish the  novel clues about the early evolution and formation  of the spiral galaxy , and its dependence on  the character of H I gas with cosmic time. The remaining paper is structured as follows: Section \ref{model} provides a comprehensive backdground  related with the redshift drift phenomenon and the models of cosmic acceleration, the observation and analysis of data are described in Section \ref{abs};  the measured results and make comparisons with previous work are delineated in Section \ref{sec:dis}; Section \ref{sec:sum} summarizes the findings derived from the HI 21cm absorption  to determine redshift drift signal.

\section{Cosmic acceleration model}\label{model}
In the comoving coordinate framework ($\rm ds^2=0$), the redshift evolution of a specific radio source over cosmic time t, the radiation emitted by the source from $t_s$ to $t_s + \Delta t_s$. It is lately observed  from $t_0$ to $t_0 + \Delta t_0$,
since $\Delta t /t \ll 1$, one can lead to the redshift drift.  
where $t_s$ is the time at which the source emitted the radiation, $t_0$ is the time at the observer. Relying on the defination of Hubble parameter $H(z) = \dot{a}/a$, then redshift drift $\dot{z} = \Delta z /\Delta t $ can be defined as:
\begin{equation}
 \rm\dot {z}   = H_0(1+z) -H(z),
\end{equation}
and where
\begin{equation}\label{eqs:hubble}
\rm H(z) = H_0\sqrt{\Omega_m(1+z)^3 + \Omega_{DE}(1+z)^{3(1+\omega)}}
\end{equation}
 is the Hubble parameter equation  with a equation of state (EoS) of dark energy $\rm \omega $ , once the contributions of cosmic radiation and  curvature flatness is assumed. $\rm H_0= 100$ hkm/s/Mpc)  is today Hubble constant. As the tracer of the expansion rate of the Universe,  $\rm \dot{z} > 0  \quad or <  0$ signifies the accelerated and decelerated expansion of the universe, respectively ,and ż = 0 meaning it  is coasting\citep{Balbi_2007,2010PhLB..692..219J}. Identically , the  spectroscopic velocity change during  an observing period $\Delta t$ for a comoving object   , $\rm \Delta v \equiv c \Delta{z}/(1+z)$, and the frequency drift  based on the of $\rm \nu_1  =  \rm\nu_{21} /(1+z) $ and  $\rm \nu_2  =  \rm\nu_{21}/(1+z+\Delta z) $  can be read:  
\begin{equation}
\rm \Delta \nu = \nu_2 -\nu_1 \approx  -\nu_{21} \frac{\Delta z}{(1+z)^2},
\end{equation}
here $\rm\nu_{21} = 1420.405751768$ MHz for the rest frame frequency of H I 21 cm absorption line and $z$ is the first measured redshift, $\nu_2$ and  $\nu_1$ is first  and second observed frequency,respectively, c is light of speed.

\begin{figure}
	\centering
	\includegraphics[width=0.7\textwidth]{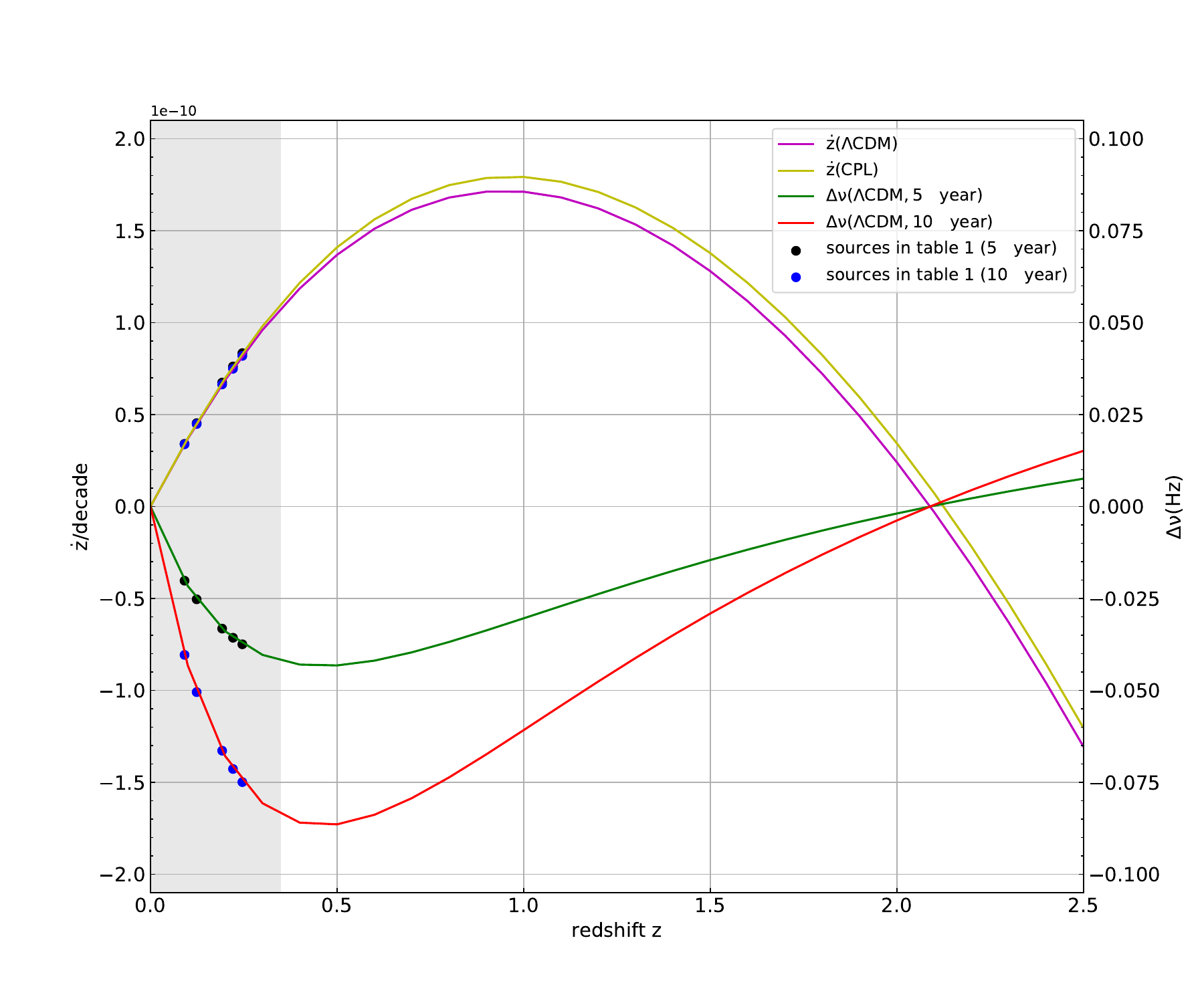}
	\caption{Theoretical   redshift drift $\dot{z}$(left) in  $\rm\Lambda$CDM(magenta) and CPL model(yellow) and the corresponding  frequency change $\Delta\nu$ (right) along with redshift  at  the observing time period of 5 year(green) and 10 year(red) under the base $\rm \Lambda CDM$, the five objective sources in table \ref{tabs:1} are marked in  blue and black  dots,respectively. The shallow gray region as the redshift scope of FAST covers.\label{figs:zh}}
\end{figure}
\begin{table*}[!htp]
	\centering
	\resizebox{\textwidth}{15mm}{
		\begin{tabular}{cccccccccc}
			\hline
			\hline
			
			backgroud source & RA(J2000) & DEC(J2000) &$z_b$& $z_{\mathrm{DLA}}$ & Type & $S_{1.4\mathrm{GHz}}$   & $\tau$ & FWHM                        & ref. \\
			& (hh mm ss) & (dd mm ss) & && & (Jy) & & ($\mathrm{km\ s^{-1}}$) & ($\mathrm{km\ s^{-1}}$)\\
			\hline
			\textbf{PKS 1413+135} & 14 15 58.818 & +13 20 23.71 &0.2467 &0.246079 & A & 1.142 &         0.34 & 18                  &      \citep{1992ApJ...400L..13C}  \\
			\textbf{B2 0738+313 A} & 07 41 10.703 & +31 12 00.23 &0.6310 &0.091235 & I & 2.051 &      0.08 & 13.4                   &      \citep{2000ApJ...532..146L}  \\
			\textbf{B2 0738+313 B} & 07 41 10.70&+31 12 00.2 & 0.6310&0.220999 & I & 2.285&              0.042 & 8                  &  \citep{Kanekar_2001}   \\

			\textbf{J094221+062334} & 09 42 21.98& +06 23 35.2 &0.1237&0.12368 & A & 0.106 &             0.723 & 30                     &     \citep{2016MNRAS.462.4197C} \\
			\textbf{J084307+453743} & 08 43 07.10 & +45 37 42.9 &0.1920 & 0.19195 & A & 0.260 &                  0.273 & 80               &   \citep{2015AA...575A..44G}  \\
			\hline
			
	\end{tabular}}
	\caption{The global information of five objective DLA systems: from first to last column denotes background source name,the absorber right Ascension(RA) and declination(DEC),the redshift of background source $\rm z_b$,the redshift of absorber $\rm\ z_{DLA}$, the type of absorber including intervening(I) and associated (A), the flux at 1.4 GHz from  VLA survey\citep{1996ADIL...JC...01C}, optical depth $\tau$ and line width  at 50\% of the absorption peak and the original literature,respectively. \label{tabs:1}}
\end{table*}

Figure \ref{figs:zh} shows the redshift drift  $\rm\dot{z}$   and the corresponding frequency changes $\rm\Delta \nu$ by the timescale of 5-year and 10-year under the   $\rm \Lambda CDM$(Planck 2018 result,P18) cosmology \citep{2020A&A...641A...6P} and Chevallier-Polarski-Linder (CPL,$\omega_0=-1$ and $\omega_a=-0.1$) parametrization \citep{2001IJMPD..10..213C,2003PhRvL..90i1301L,2007MNRAS.382.1623B} in colorful lines ,respectively,and the five objective sources in table \ref{tabs:1} are marked in solid dots  that will be observed by FAST for redshift drift experiment. Based on this figure, dark energy candidates can be effectively constrained at high redshifts using SL signal. However, distinguishing  the two mainstream candidates becomes challenging at low redshifts without combining various observational data. To achieve a frequency resolution of less than 0.1 Hz for the five candidate sources listed in table \ref{tabs:1} within the shaded area of figure \ref{figs:zh} , the observing period needs to be extended to 10 or 20 years or even longer, along with the development of more advanced backend techniques to improve spectral accuracy.   

\begin{figure}
	\centering
	\includegraphics[width=0.7\textwidth]{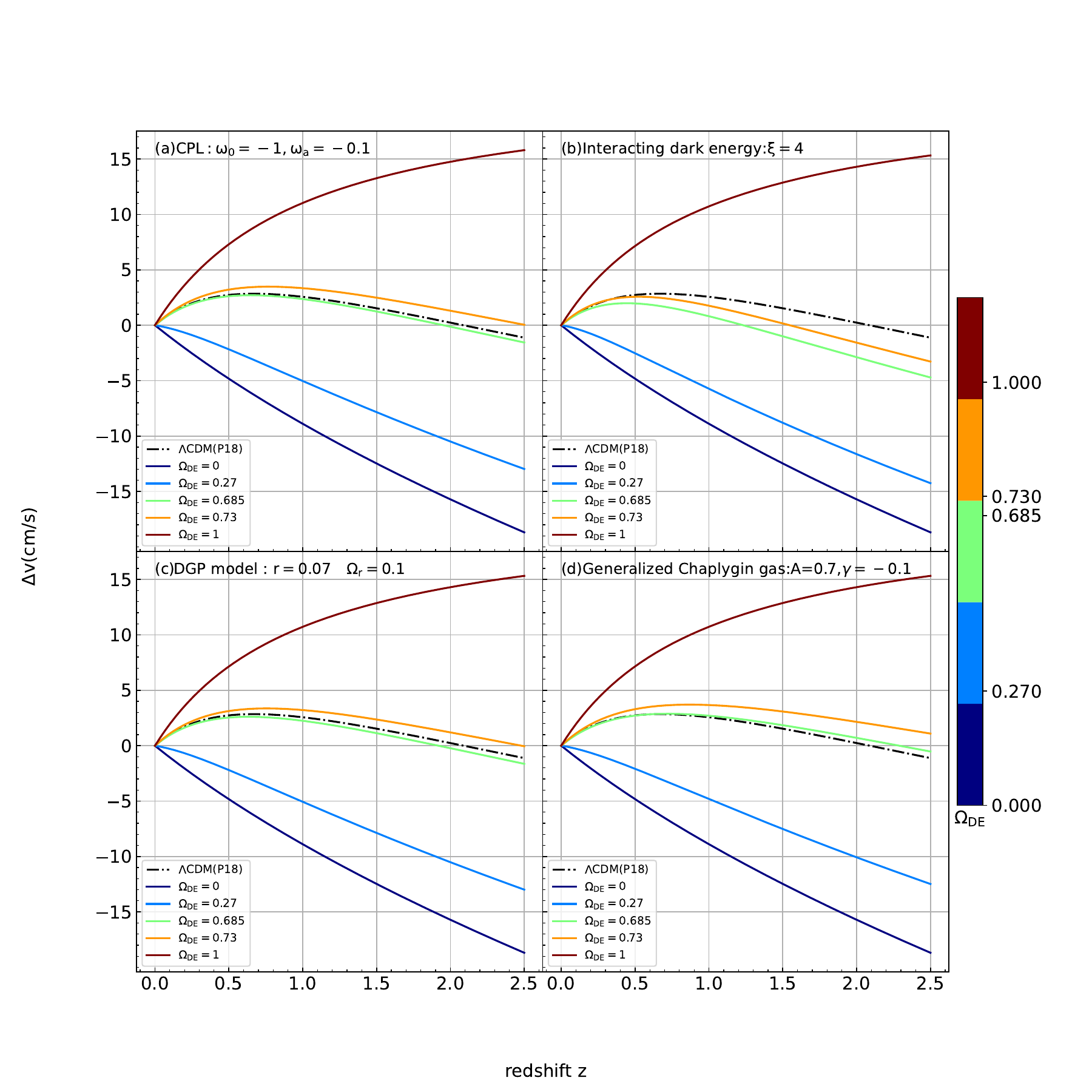}
	\caption{The  evolution  of velocity drift $\rm\Delta v$ as a function of  redshift when fixed the observing time-span $\Delta t$ =10 yr(year) against a same object under the four alternative cosmological models. The each colorful curve denotes a certain value of $\rm\Omega_{DE}$  as the ticks on right colorbar, the black dashdotted stands for the $\rm\Delta v$ under the $\rm \Lambda CDM$ model as the reference case.\label{figs:vv}}
\end{figure}

Alternatively, the redshift drift effect can be quantized using the velocity drift $\rm\dot{v}$, typically a few mm/s per year. This effect is debated in alternative cosmological models with specific model parameters, highlighting the role of dark energy with energy density $\rm\Omega_{DE}$ ranging from 0 to 1 in the redshift volume $z < 2.5$ as the results in figure \ref{figs:vv}. The figure shows four competing cosmological models used to predict the evolution of velocity drift $\rm \dot{v}$ with redshift z, based on changes in dark energy density $\rm\Omega_{DE}$ from 0 to 1 over a 10-year observation period with a fixed $\rm H_0 =70$ km/s. Further investigation into redshift drift effects under different cosmological scenarios can be found in previous researches  \citep{Balbi_2007,2008MNRAS.386.1192L,2021MNRAS.508L..53E}. The behavior of each model of plotted $\rm\dot{v}$ from (a) to (d),  $\Lambda$CDM  framework as a reference, which is described in central black dash dotted line in each panel fixing $\rm\Omega_{DE}$=0.685. The variation of $\rm\Omega_{DE}$ ranges from 0.685 to 0.73 across panels (a-d), the trajectory of $\rm\Delta v$ in the four models mimics that of $\rm\Lambda$CDM. By observing the velocity drift $\rm\dot v$, the alternative cosmological models  can be further  differentiated , especially at high redshifts z$>$0.5 as results in plot (a) to (d). For $\rm\Omega_{DE}$ =0.685 case, the $\rm\Delta v$ in (a),(c) and (d) are completely common tendency compared with  in $\rm\Lambda$CDM  at redshift z$\le$ 0.5, and $\rm\Delta v$ in (b) somewhat less 0 to 1 cm/s than in $\Lambda$CDM  from redshift z= 0 to 0.5. However, in  the case of $\rm\Omega_{DE}$ =0.73, at redshift z$\le$ 0.5, the prediction of $\rm\Delta v$ in (b) mostly agrees with the  $\rm\Lambda$CDM , otherwise (a),(c) and (d) uniformly larger than 0 to 1 cm/s,  the few differences in those models cannot be the elements of the specific free parameters since have been fixed, exactly the biases among these  models are about to be carefully discussed with the advent of the  high precision data , such as  SKA and ETL. Owing to the arrange we propose, first 5  targeted sources in table \ref{tabs:1} will be observed, and  the range of S/N can lie in  50 to 1000 as the current status of FAST,the experiment time span as the   periods  of $\Delta t$ = 5 and 10 or 20  years and imposing exposure time on source. The appreciable amount of DLA systems surely reduce the systematic error of the spectral line  and enhance the measured accuracy to reinforce  the reliability of the experiment.

\section{ Observation  and analysis  }\label{abs}

\begin{figure}[ht]
	\centering
	\includegraphics[width=.7\textwidth]{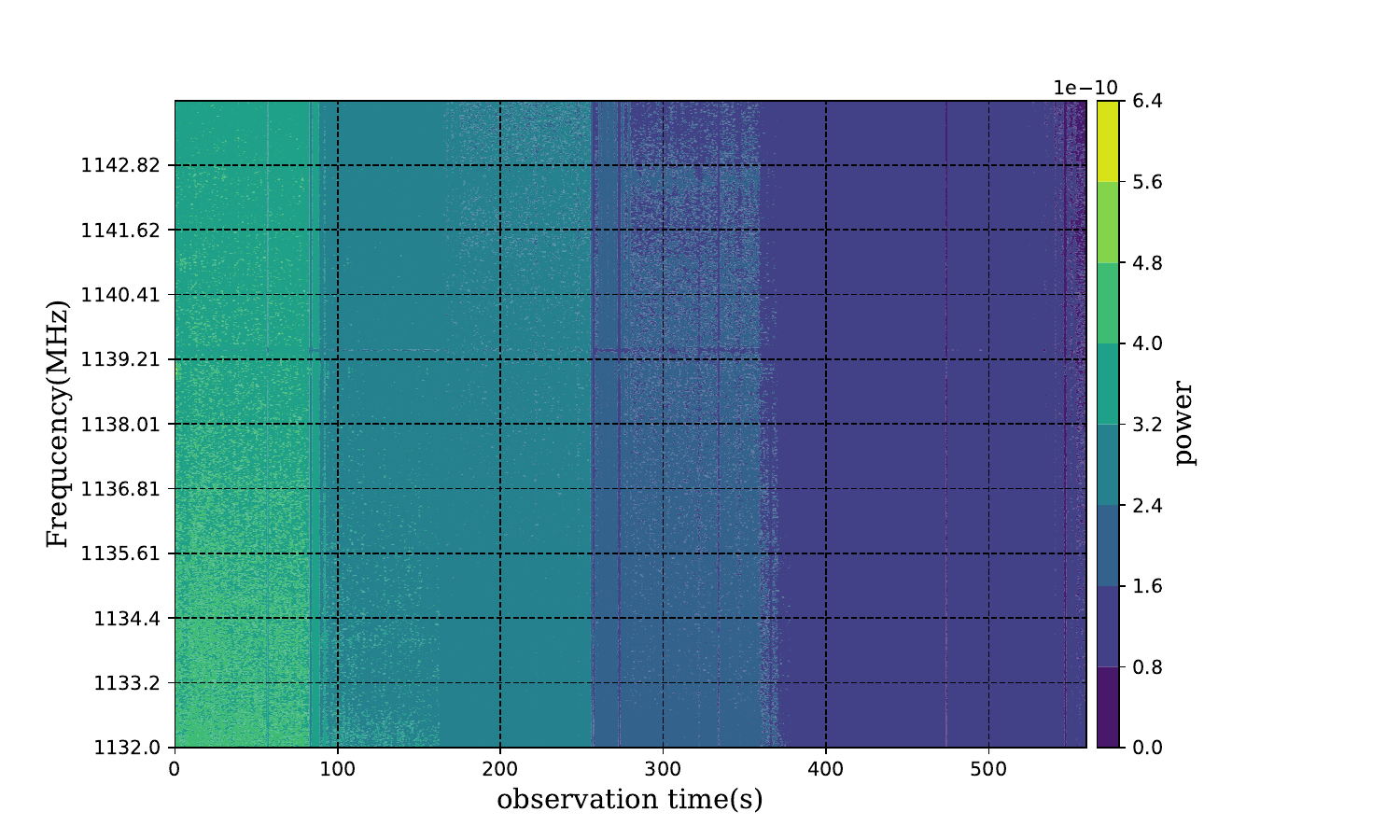}
	\includegraphics[width=1\textwidth]{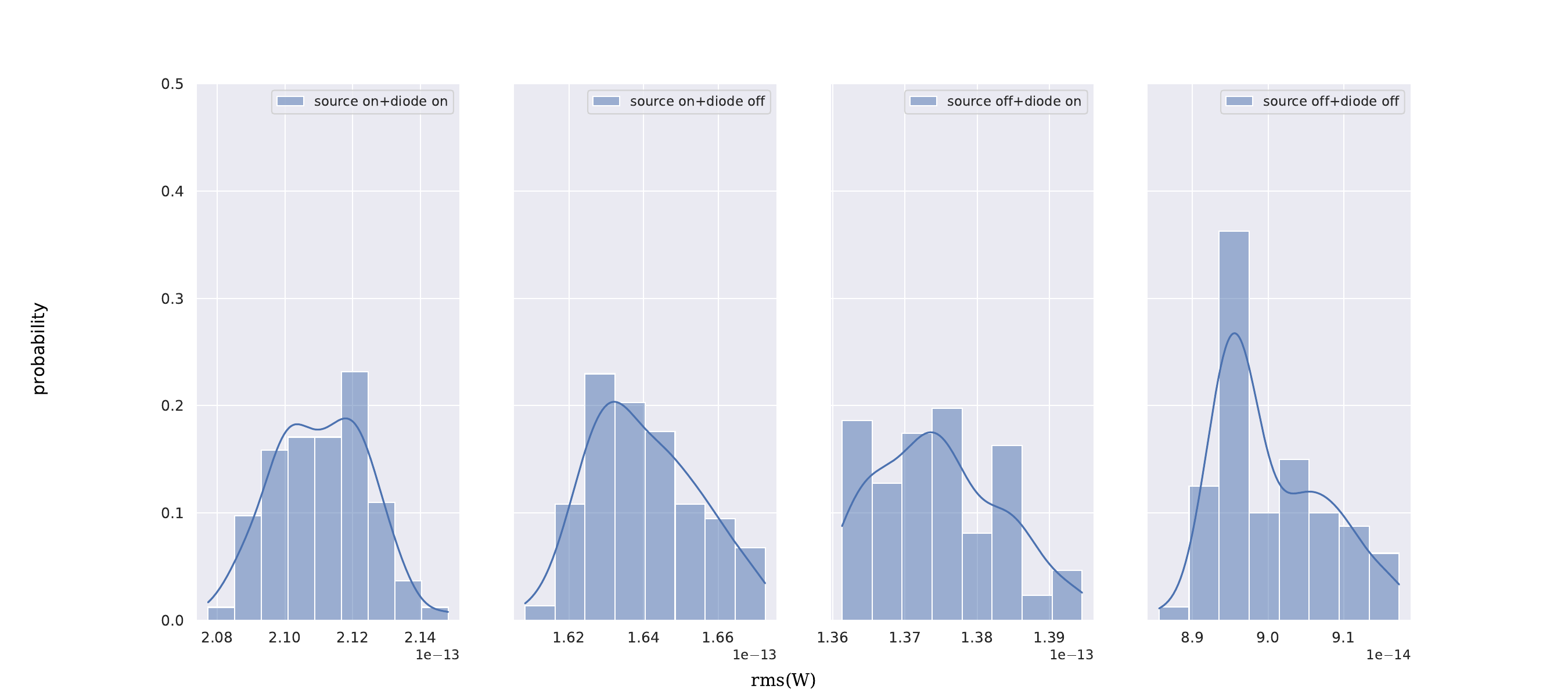}
	\caption{Top panel: the power intensity  of the the absortption line in background spectrum of PKS1413+135   along with the different effective observational time of the four different observational status: source-ON/OFF and diode-on/off.  A faint absorption line near 1139.21 MHz in the horizonal direction can be seen. Below histogram shows the probability  distribution of the root mean square(rms)  level of the data in raw data power,solid curve  fits trendency of data distribution.}
	\label{figs:im1}
	
\end{figure}
\begin{figure}[th]
	\centering
	\includegraphics[width=.7\textwidth]{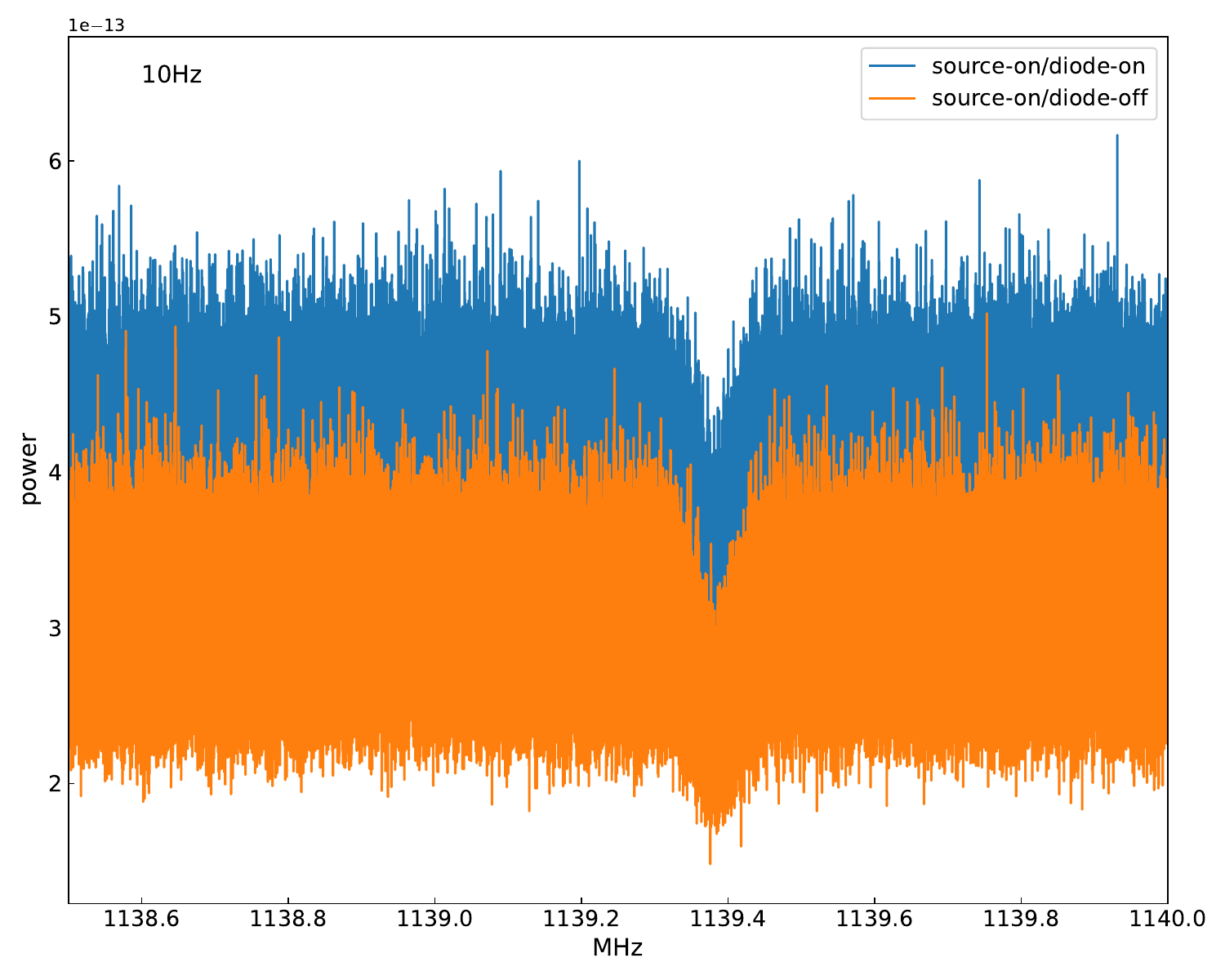}
	\caption{The power of the raw data(the PKS1413+135 source) with the raw data frequency resolution of 10Hz with source-on-diode-on and source-on-diode-off mode. The absorption  feature clear seen near 1139.21 MHz.}
	\label{figs:im2}
\end{figure}
 
FAST conducted a 10-minute observation of the DLA system at redshift $z$ = 0.24670045 toward PKS1413+135 using its 19-beam receiver in L-band to sample data from two polarizations at the central beam in ON-OFF mode (300 seconds in ON mode).  By adding 10 K noise, the raw spectrum is obtained using the fast Fourier transform (FFT) technique, with a cut-off frequency range of 1132-1144 MHz and stored at a resolution of 10 Hz.

 The top part of figure \ref{figs:im1} shows the temperature intensity of PKS1413+135 in Kelvins (K) at 1139.21 MHz, indicating a shallow absorption line with the temperature around 16-17 K. About 322 seconds data was effective  out of a total of 636 seconds, with the remaining 314 seconds data showing unstable  possibly due to instrument effects. The histogram below figure \ref{figs:im1} displays the probability distribution of each root mean square (rms) bin of raw data, illustrating four modes when outside the absorption line and RFI band.The rms level distribution on-source basically follows a Gaussian profile, with weighted rms values of 2.103 $\times 10^{-13}$  and 1.644 $\times 10^{-13}$  . While the rms values under source-off mode are relatively diffuse, with the weighted rms being 1.371 $\times 10^{-13}$  and 9.06 $\times 10^{-14}$ , respectively, and  the final S/N can be estimated based on the four rms level states.  Figure \ref{figs:im2} shows absorption lines at 1139.21 MHz with a frequency resolution of 10 Hz in source-on/diode-on and source-on/diode-off modes.

To calibrate the raw power to temperature and then to flux, the 322 -second effective data. The 154 seconds with the source on and 168 seconds with the source off, can be processed in the following manner to convert the raw data into temperature as following:
\begin{equation}\label{eqs:T}
\begin{aligned}
\rm
T_1=&T_\mathrm{sys}^\mathrm{on}=\frac{P^\mathrm{on}}{P^\mathrm{off}-P^\mathrm{off}}T_\mathrm{cal}-T_\mathrm{cal},\\
\rm  
T_2= &T_\mathrm{sys}^\mathrm{off}=\frac{P^\mathrm{off}}{P^\mathrm{on}-P^\mathrm{off}}T_\mathrm{cal},
\end{aligned}
\end{equation}

\begin{equation}\label{eqs:w}
 w_i =\sigma^2_{j}/(\sigma^2_1+\sigma^2_2),\quad i,j =1,2, i\ne j,\quad   
 T_\mathrm{sys}= w_1T_1 +w_2T_2.\\
		\rm  
\end{equation}

\begin{figure}[tbp]
	\centering
	\includegraphics[width=.6\textwidth]{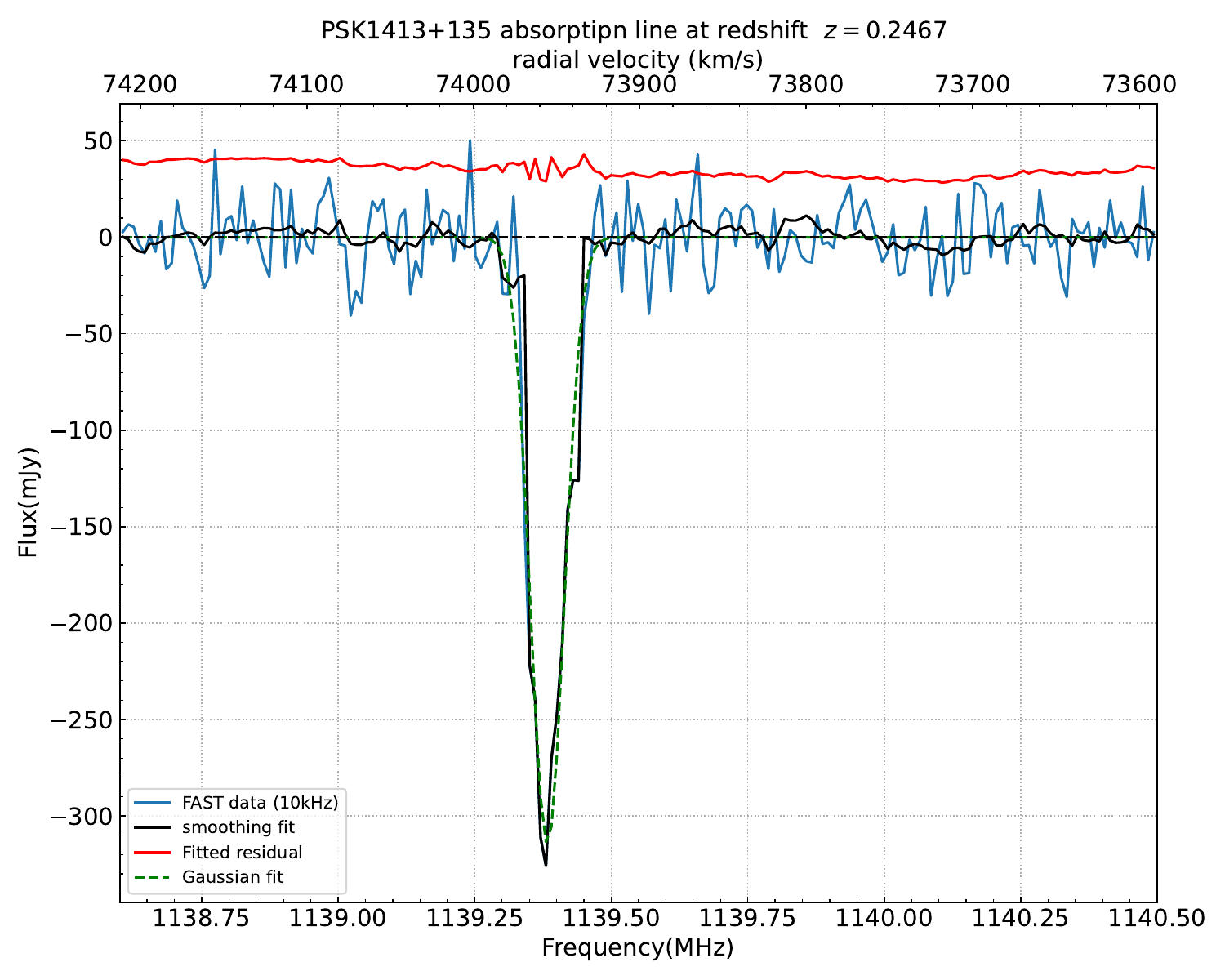}
	\caption{The H I absorption feature of PKS1413+135 in the spectral resolution of 10 kHz. The light blue line represents the raw data, while the black curve is a smoothed line within a 0.5 MHz bandwidth. The green dotted line represents a single Gaussian function profile, while the red solid line shows the residual between the original data and the fit, approximately 40 mJy.}
	\label{figs:10k}
\end{figure}

In the equation, $P$ represents the measured power , and $T_\mathrm{cal}$ refers to the temperature of the noise diode, which is assumed to be 10 kelvins (K), $\rm\sigma_1$ and $\sigma_{2}$ stands for the  rms levels for source-on and source-off,respectively. To smooth out standing waves and baseline fluctuations, apply a gaussian convolution function with a 3$\sigma$ width, except  the region of the absorption peak. Then, to correct the raw spectrum baseline by fitting a combined polynomial and sine function, and remove this fitted baseline from the original dataset. We analyze the absorption feature by fitting it with a single Gaussian function to determine the optical depth at the peak. Lastly, we calculate the flux density in millijanskys (mJy) by dividing with the antenna gain factor of 16.48 Kelvin per Jansky (K/Jy), as demonstrated in the figure \ref{figs:10k}, with a frequency resolution of 10 kHz. The equation below is used to compute the column density ($\rm N_{H I}$) and the optical depth ($\tau$) of the DLA system\citep{2021MNRAS.503.5385Z}: 
\begin{equation}
\rm N_{H I} = 1.823 \times \frac{T_s}{f}\int\tau dv, \quad \\ 
\rm \tau = - ln\left( 1+ \frac{S_{H I}}{S_{1.42G}}\right),
\end{equation}
In this context, $\rm T_s$ represents the H I gas spin temperature, f indicates the absorber's covering factor, which typically defaults to one, and v is the radial velocity  in km/s. The $\rm S_{H I}$  represents the intensity of the absorption line, indicated by a negative number, while $\rm S_{1.4GHz}$ denotes the strength of the radio emission at 1420.4 MHz as recorded by the NVSS survey\footnote{https://www.cv.nrao.edu/nvss/NVSSlist.shtml}.
The signal-to-noise ratio (S/N) is calculated as the ratio of peak flux to the rms level \citep{Kang_2022}: 
 \begin{equation} 
\rm S/N=\frac{F_p}{\sigma(F)},\quad \sigma(F) = ( \sigma_{rms,2}  +\sigma_{rms,4})/4
\end{equation}
The F represents peak flux, while $\rm \sigma (F)$ indicates the mean rms level (mJy) of the absorption line only considered  in diode-off cases. $\rm \sigma_{rms,1}$  to $\rm \sigma_{rms,4}$ are computed by the mean of four case weighted rms, $\rm\sigma_i = \Sigma P_i  \times rms_i $ as the histogram of figure \ref{figs:im1}, $P_i$ labels the probability and its corresponding $rms_i$  in each bin. To correct the heliocentric radial velocity in Astropy's coordinate module, the observed radial velocity in radio frequency ($v_m = c (\nu_0 -\nu_m)/\nu_0$) needs to be added after subtracting the baseline of the spectrum as a linear approximation\citep{thomas_robitaille_2023_7963327}, then $\rm v$ can be displayed as following\citep{Jiao_2020}:
 \begin{equation}\label{eqs:v2}
 \rm v =v_m +v_b + \frac{v_m v_b }{c} \quad    (\mathrm{km\ s^{-1}}),
 \end{equation}
 In figure \ref{figs:10k}, v represents the final radial velocity on the x-axis. $\rm \nu_0$ and $\nu_m$ are the rest and observed frequencies of the 21 cm line, respectively. $\rm v_b$ is the barycentric correction velocity based on the telescope's position on Earth at a specific observation time, and c is the speed of light in vacuum. The barycentric correction velocity $\rm v_b$ for the absorption line of PKS1413+135 is 1.33  km/s, with an accuracy of about 3 m/s for the source. The peak position of the absorption profile indicates the redshift drift signal of cosmic acceleration, $\sigma_{\mathrm{V}} $ represents the profile's narrowness, a common method validated for single-dish FAST data \citep{2004AJ....128...16K,2021MNRAS.503.5385Z}:
 \begin{equation}\label{eqs:v3}
 	\sigma_{\mathrm{V}}=3\frac{\sqrt{PR}}{S}\ (\mathrm{km\ s^{-1}}),
 \end{equation}
 
$R$ represents velocity resolution, $S$ stands for S/N, $w_{20}$ and $w_{50}$ represent the width of the absorption line at 20\% and 50\% levels, respectively. The parameter $\rm P=(w_{50}-w_{20})/2$ quantifies the sharpness of the profile edges\citep{2004AJ....128...16K,2021MNRAS.503.5385Z}.The uncertainty estimation is applied in the analysis of frequency, redshift, and velocity in section \ref{sec:dis}. 
 
\section{\rm Results and Discussions}\label{sec:dis}
The absorber was exposed about 156 seconds with a frequency resolution of 10 Hz. The data was  processed to show radial velocity from 73940 to 73970 km/s using a 10 kHz frequency resolution and the bandwidth was limited to 2 MHz as shown in figure \ref{figs:10k}. We use a single Gaussian model to fit the absorption profile, with the best-fit parameter values shown in Table \ref{tabs:2}. The total flux density obtained is 1053.72 $\pm$ 37 mJy, with a peak value of 338.016$\pm$4.419 mJy. The column density is $\rm N_{H I} = 2.2867 \rm \times 10^{22} cm^{-2}$, and the integrated optical depth at the peak is $\tau = 0.329 \pm 0.021$, assuming a spin temperature of $\rm T_s = 300 $ K in a spiral host galaxy. The result is in agreement with the form of $\rm N_{H I} \simeq 1.3 \times 10^{19} \times T_s/f$ $\rm $ $\rm$ $\rm cm^{-2}$, given that $f \le 0.1$, indicating that the source has been well resolved \citep{1992ApJ...400L..13C}. 

\begin{table}[tbp]\scriptsize
	\centering
	\begin{tabular}{|c|c|c|c|c|c|c|c|c|}
	    \hline
		\hline
		Band width & S/N & $\mathrm{res_\mathrm{\nu}}$ & $\mathrm{res_v}$ & $\mu$ & $\sigma$ &a   & $\sigma_{\mathrm{V}}$ \\
		(MHz)           &         & (kHz) & ($\mathrm{km\ s^{-1}}$) & ($\rm km/s$) & ($\mathrm{km\ s^{-1}}$) & (mJy) &  (km/s) \\
		 \hline
		1138.0$\sim$1140.0 & 57.4357 & 10 & 3.284  &  73958.9408& 9.3071& -55.483 & 0.2306\\   
		\hline
	\end{tabular}
	\caption{The parameters used to fit the DLA in PKS1413+135 spectra within a 2 MHz bandwidth are all best-fit values assumed to follow a Gaussian profile $f(x)=a\exp{\frac{(x-\mu)^2}{2\sigma^2}}$. The final columns indicates the velocity uncertainty $\sigma_{\mathrm{V}}$.\label{tabs:2}}	
\end{table}

The H I 21 cm absorption line profile in DLA systems against background radio sources shows complex spectral structures with multiple asymmetric components, reflecting the presence of unsettled gas in the ISM. This includes gas outflows from radio jets or tidal streams, as well as infalling H I gas towards central black holes in host galaxies \citep{2013MNRAS.436.2366R,2018A&ARv..26....4M,2023A&A...675A..40H,2015A&A...575A..44G,2001A&A...373..394K},and the minor wing components are needed for precise fits, yet they are omitted from the $\dot{z}$ measurement when the signal-to-noise ratio is sufficiently high\citep{1992ApJ...400L..13C,Darling_2012}.The DLA system facing PKS1315+145 features two Gaussian components within its absorption line at z = 0.24670374, according to the analysis\citep{Darling_2012}. Due to the brief 10-minute observation period of this source, we were unable to clearly identify all the minute components in our analysis. As a result, it was not possible to resolve the finer structures in detail, and we had to represent the entire spectral line with a simplified Gaussian profile, which includes a significant damping wing, as depicted in the figure \ref{figs:10k}. The observed data aligns precisely with the Gaussian fit, resulting in a total flux of $\rm F = 1045.747 mJy \pm 0.192$. This provides a signal-to-noise ratio (S/N) of 57.4357 at a frequency resolution of $\rm\Delta \nu =10 $ kHz. Additionally, the barycentric radial velocity is determined to be $v = 73944.3109\pm 0.1139$ km/s, with a redshift ($z$) of $0.24670021\pm 0.0011$ and a central frequency ($f$) of $1139.39017\pm 0.1346$ MHz. The findings align with the prior study, acknowledging some uncertainty\citep{1992ApJ...400L..13C}.

\begin{figure}[tbp]
	\centering
	\includegraphics[width=0.48\textwidth]{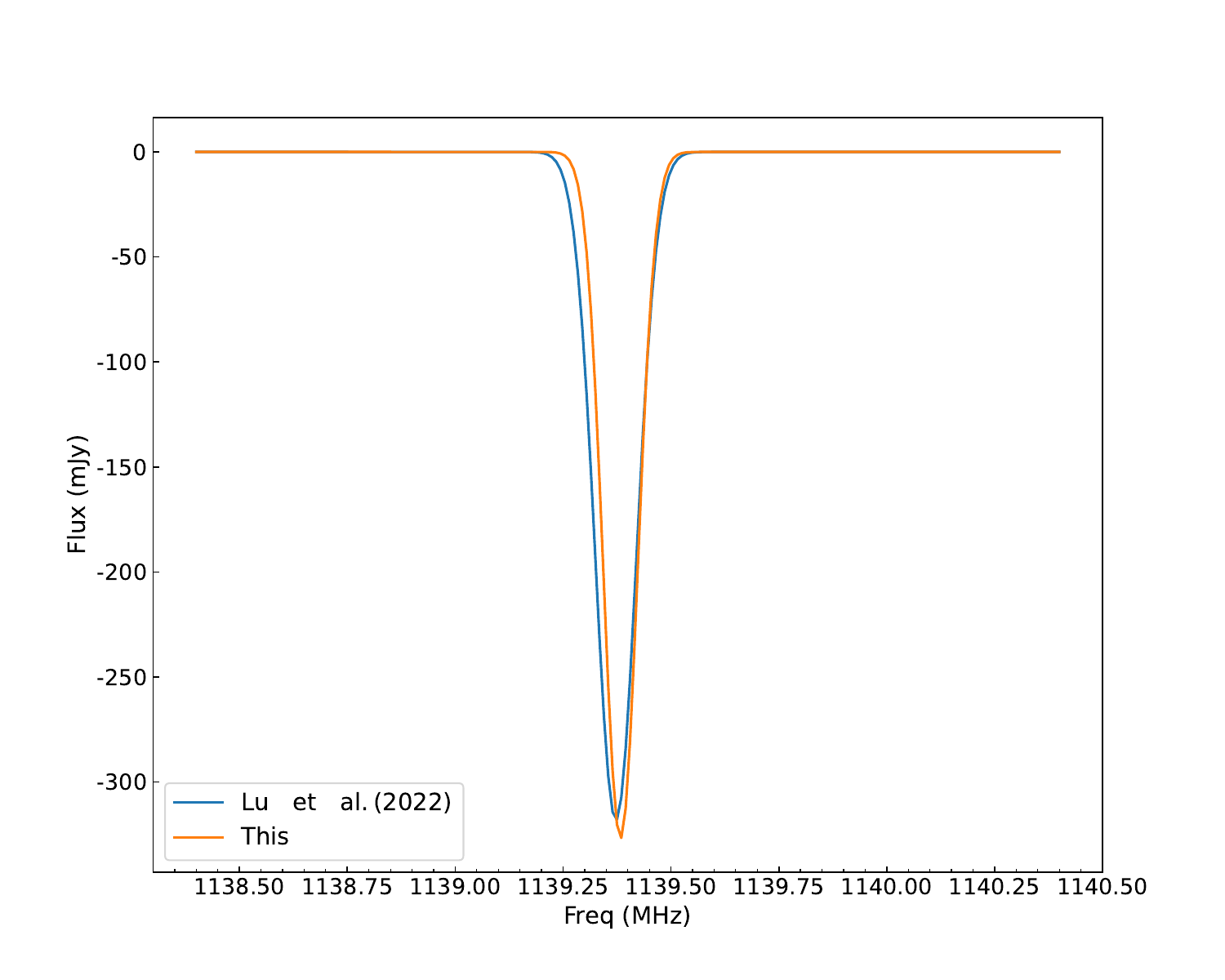}
	\includegraphics[width=0.48\textwidth]{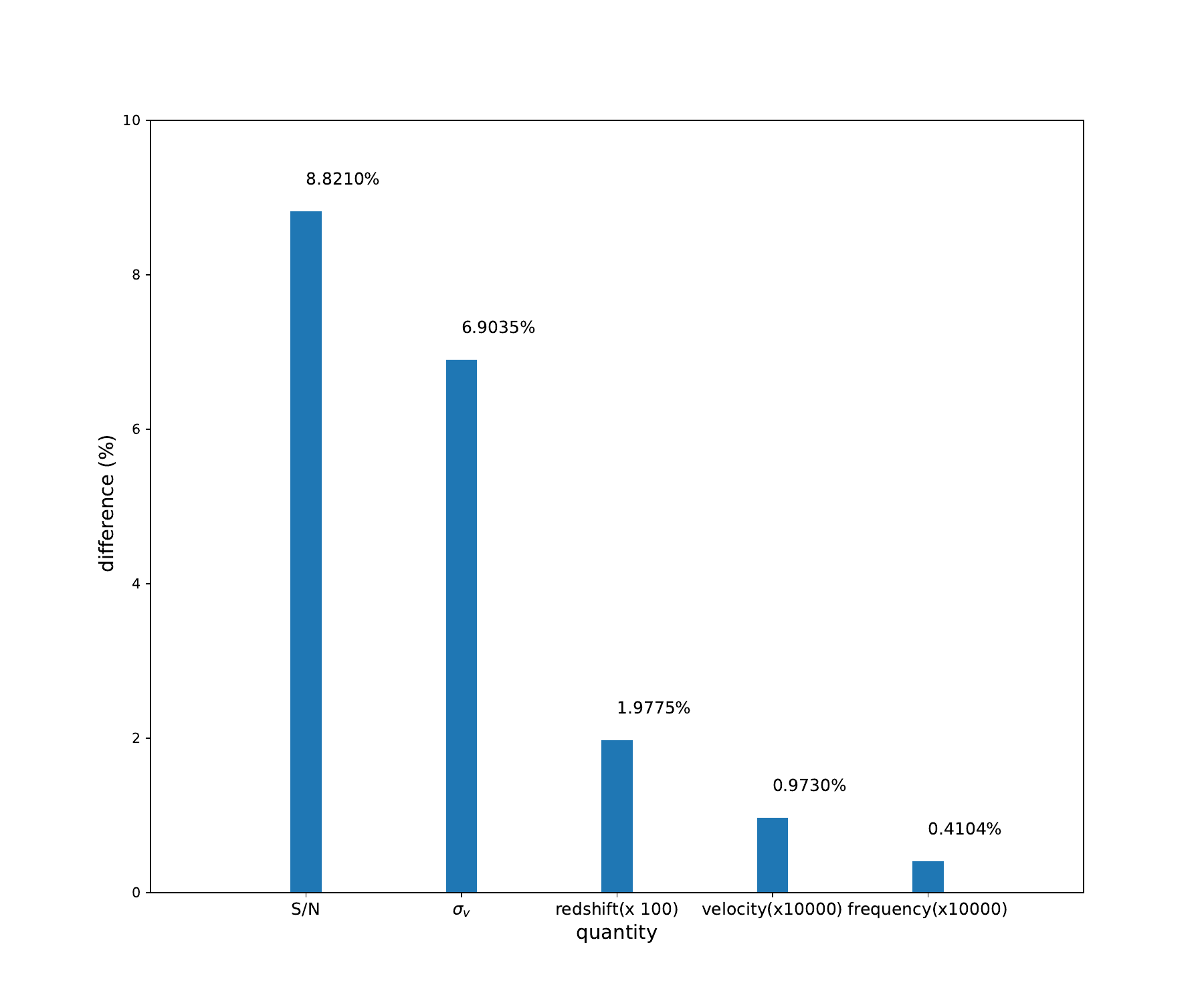}
	\caption{Left panel: The fitting comparsion of H I absorption feature of PKS1413+135 in this work with L'result of \citep{2022PDU....3701088L},in blue and orange line respectively. Right panel: the relative errors are presented in bar  in percence form of S/N, velocity error,redshift,frequency,total flux between two works,This work shows the results of the data process.}
	\label{figs:error1}
\end{figure}

In reality, the data of the absorption lines has been examined as preparatory work \citep{2022PDU....3701088L}.The line is resolved with frequency resolutions of 0.1, 1, and 10 kHz in three bands: 1135.0-1145.0 MHz, 1137.4-1141.4 MHz, and 1138.4-1140.4 MHz, respectively. A single Gaussian profile is used to fit the absorption feature, resulting in various signal-to-noise ratios ranging from 8.08 to 55.43. The velocity uncertainty $\sigma_{\mathrm{V}}$ is at the level of 0.11 km/s. In contrast, we revisit the raw data and process it using an alternative approach represented by the equation \ref{eqs:T} related to the H I emission method of extragalaxies. The line is modeled with a Gaussian profile as indicated in table \ref{tabs:2}, with the fitting results displayed in the left panel of figure \ref{figs:error1}. In Lu's study, the absorption peak is at (1139.375, -318.3) on the blue line, while in this study, it's at (1139.383, -326.9) on the orange line. The frequency has shifted by approximately 0.005 MHz and there's an 8 mJy difference in peak flux. Other quantities such as signal-to-noise ratio (S/N), velocity error, redshift error, and the total flux of the absorber can also be calculated as their relative errors using equation \ref{eqs:er2} as shown in the right panel. The error bars indicate a consistent decrease in error from left to right, with the error remaining below 10 percent. The signal-to-noise ratio has the highest error at 8.821\%, while the frequency has the least at just 0.0000004\%. For clarity in the graphical representation, the error percentages for velocity and frequency have been amplified by a factor of 10,000, and for redshift by a factor of 100.
\begin{equation}\label{eqs:er2}
	\rm\ difference (\%) = \frac{Q_T -Q_L}{Q_L}
\end{equation}
The equation presents the relative error for these measurements at the percent level, with T denoting this study and L indicating the results from \citep{2022PDU....3701088L}. The comparative differences can be seen in the chart referenced as figure \ref{figs:error1}.
  In addition, as spectral resolution increases, the velocity uncertainty $\sigma_{\mathrm{V}}$ diminishes, leading to more accurate results due to fewer geometric distortions and more thorough processing of the initial data, according to equation \ref{eqs:v3}, does nothing with the depth of the absorption line peak. Despite limited observation time of the source, the baseline fluctuations and various spectral RFI noises have been successfully removed using Gaussian fitting and  RFI mitigation algorithm, allowing for the clear extraction of the absorption line's global information, as shown in figure \ref{figs:10k}. The observed radial velocity has been corrected to the heliocentric frame by applying a barycentric adjustment of 1.33 km/s using the astropy method.
 
Accurately detecting the tiny redshift drift, as the indicator of cosmic acceleration,  and to establish a definitive SL cosmology, it's essential to improve data accuracy. This includes meeting spectral resolution below 0.1 Hz and velocity measurement accuracy on the order of mm/s, requirements comparable to those of the SKA or ELT, along with compiling a comprehensive catalog of DLA systems and moreover, distinguish the unique acceleration-related pollution caused by nearby systems. FAST's spectral resolution for the SETI experiment the expected frequency accuracy of 3.725 Hz,actually  achieved 7.5 Hz per channel \citep{2020ApJ...891..174Z,2022AJ....164..160T}. In this case  an efficient approach to increase the detection of redshift drift signal is to extend the observation period of H I 21 cm absorption lines in DLA systems against the same background source. The recent release of the extensive survey of H I emission from extragalactic objects by FAST surely includes host galaxies of potential DLA candidates. These DLA system catalogs continuously either improve  accuracy or relieve error of redshift drift signal via the data from H I 21 cm absorption in nearby cosmological volumes\citep{Kang_2022,2023arXiv231206097Z}.

\section{Summary} \label{sec:sum}
  Initially, we present the concept and framework of redshift drift as a Cosmic Accelerometer through the background of cosmological dynamics. The analysis in figure \ref{figs:zh} shows that a radio telescope must have a spectral resolution of less than 0.1 Hz over a 5 or 10 year observation period to accurately detect and measure the rate of cosmic acceleration. In our next proposal, we will  observe  the five objective sources at various redshifts and relative informations can be found in figure \ref{figs:zh}. Furthermore, the predicted spectroscopic velocity drift $\rm\dot{v}$ is analyzed for four alternative cosmological frameworks beyond the $\Lambda$CDM model by varying the dark energy density $\rm\Omega_{DE}$ from 0 to 1. These frameworks include CPL, Interacting dark energy, DGP model, and Generalized Chaplygin gas. These models can be distinguished easily at high redshifts ($z > 0.5$),  showing pretty small differences of 0-1 cm/s at lower redshifts.  Utilizing high-quality redshift drift data and other cosmic probes at different scales to further research is needed. Our study centers on examining the H I 21 cm absorption line in the DLA system linked to PKS1413+135, observed at a redshift of $z = 0.24670041$. This observation, conducted by FAST with a spectral resolution of 10 Hz, lasted for 10 minutes. The final measurements indicate a radial velocity of $v = 73944.3109 \pm 0.1139$ km/s at the absorption line's peak, with a redshift of $z = 0.24670021 \pm 0.0011$. The central frequency is calculated to be $f = 1139.39017 \pm 0.1346$ MHz. The neutral hydrogen column density, derived from the integrated optical depth $\tau = 0.329 \pm 0.021$, is $\rm N_{H I} = 2.2867 \times 10^{22}   /cm^2$, assuming a spin temperature of 300 K as a spiral host galaxy. This value aligns with the expected column density formula $\rm N_{H I} = 1.3 \times 10^{19}\times (T_s/f)   /cm^2$ when considering the spin temperature's uncertainty.

 The comparison of the absorption line data with the study of Lu\citep{2022PDU....3701088L} is depicted in figure \ref{figs:error1}. The left panel shows the Gaussian fit, while the right panel presents the relative errors in signal-to-noise ratio (S/N), velocity dispersion ($\sigma_{\mathrm{V}}$), measured redshift, velocity, and frequency. The relative errors are all below 10 percent, with some showing negligible discrepancies. The approach for converting raw power data into temperature, as detailed in equation \ref{eqs:T},  it  yields essentially consistent results across the two different analyses. The findings confirm that using the H I 21 cm absorption line as a probe for redshift drift effect is effective for measuring cosmic acceleration rates beyond quasar lyman-$\alpha$ forests. Efforts are underway to enhance the signal's accuracy by analyzing the overall shape of the absorption profiles, applying a Gaussian function for better fit, and identifying finer spectral details to clarify the conditions and components of H I gas in host galaxy. Over a decade, by compiling over 2000 candidate DLA systems observed across various parts of the sky with FAST, SKA or ELT, we can significantly enhance the precision of the SL effect. These improvement can reach up to two orders of magnitude and achieve the required mm/s or less than 0.1 Hz precision, thus allowing us to ultimately assert the acceleration of the universe and deepen our understanding of dark energy, using the statistical analysis of H I 21 cm absorption lines and potentially to refine real-time cosmology.
\section{Acknowledgements}\label{ac}
 This  work is supported by the National SKA Program of China(2022SKA0110202) and National Natural Science Foundation of China (Grants No.11929301). This work made use of the data from FAST (Five-hundred-meter Aperture Spherical radio Telescope). FAST is a Chinese national mega-science facility, operated by National Astronomical Observatories, Chinese Academy of Sciences.
  .

\bibliography{abs}
\bibliographystyle{raa}
\end{document}